

\font\bigbold=cmbx12
\font\ninerm=cmr9      \font\eightrm=cmr8    \font\sixrm=cmr6
\font\fiverm=cmr5
\font\ninebf=cmbx9     \font\eightbf=cmbx8   \font\sixbf=cmbx6
\font\fivebf=cmbx5
\font\ninei=cmmi9      \skewchar\ninei='177  \font\eighti=cmmi8
\skewchar\eighti='177  \font\sixi=cmmi6      \skewchar\sixi='177
\font\fivei=cmmi5
\font\ninesy=cmsy9     \skewchar\ninesy='60  \font\eightsy=cmsy8
\skewchar\eightsy='60  \font\sixsy=cmsy6     \skewchar\sixsy='60
\font\fivesy=cmsy5     \font\nineit=cmti9    \font\eightit=cmti8
\font\ninesl=cmsl9     \font\eightsl=cmsl8
\font\ninett=cmtt9     \font\eighttt=cmtt8
\font\tenfrak=eufm10   \font\ninefrak=eufm9  \font\eightfrak=eufm8
\font\sevenfrak=eufm7  \font\fivefrak=eufm5
\font\tenbb=msbm10     \font\ninebb=msbm9    \font\eightbb=msbm8
\font\sevenbb=msbm7    \font\fivebb=msbm5
\font\tenssf=cmss10    \font\ninessf=cmss9   \font\eightssf=cmss8
\font\tensmc=cmcsc10

\newfam\bbfam   \textfont\bbfam=\tenbb \scriptfont\bbfam=\sevenbb
\scriptscriptfont\bbfam=\fivebb  \def\Bbb{\fam\bbfam}
\newfam\frakfam  \textfont\frakfam=\tenfrak \scriptfont\frakfam=%
\sevenfrak \scriptscriptfont\frakfam=\fivefrak  \def\frak{\fam\frakfam}
\newfam\ssffam  \textfont\ssffam=\tenssf \scriptfont\ssffam=\ninessf
\scriptscriptfont\ssffam=\eightssf  
\def\smc{\tensmc}

\def\eightpoint{\textfont0=\eightrm \scriptfont0=\sixrm
\scriptscriptfont0=\fiverm  \def\rm{\fam0\eightrm}%
\textfont1=\eighti \scriptfont1=\sixi \scriptscriptfont1=\fivei
\def\oldstyle{\fam1\eighti}\textfont2=\eightsy
\scriptfont2=\sixsy \scriptscriptfont2=\fivesy
\textfont\itfam=\eightit         \def\it{\fam\itfam\eightit}%
\textfont\slfam=\eightsl         \def\sl{\fam\slfam\eightsl}%
\textfont\ttfam=\eighttt         \def\tt{\fam\ttfam\eighttt}%
\textfont\frakfam=\eightfrak     \def\frak{\fam\frakfam\eightfrak}%
\textfont\bbfam=\eightbb         \def\Bbb{\fam\bbfam\eightbb}%
\textfont\bffam=\eightbf         \scriptfont\bffam=\sixbf
\scriptscriptfont\bffam=\fivebf  \def\bf{\fam\bffam\eightbf}%
\abovedisplayskip=9pt plus 2pt minus 6pt   \belowdisplayskip=%
\abovedisplayskip  \abovedisplayshortskip=0pt plus 2pt
\belowdisplayshortskip=5pt plus2pt minus 3pt  \smallskipamount=%
2pt plus 1pt minus 1pt  \medskipamount=4pt plus 2pt minus 2pt
\bigskipamount=9pt plus4pt minus 4pt  \setbox\strutbox=%
\hbox{\vrule height 7pt depth 2pt width 0pt}%
\normalbaselineskip=9pt \normalbaselines \rm}

\def\ninepoint{\textfont0=\ninerm \scriptfont0=\sixrm
\scriptscriptfont0=\fiverm  \def\rm{\fam0\ninerm}\textfont1=\ninei
\scriptfont1=\sixi \scriptscriptfont1=\fivei \def\oldstyle%
{\fam1\ninei}\textfont2=\ninesy \scriptfont2=\sixsy
\scriptscriptfont2=\fivesy
\textfont\itfam=\nineit          \def\it{\fam\itfam\nineit}%
\textfont\slfam=\ninesl          \def\sl{\fam\slfam\ninesl}%
\textfont\ttfam=\ninett          \def\tt{\fam\ttfam\ninett}%
\textfont\frakfam=\ninefrak      \def\frak{\fam\frakfam\ninefrak}%
\textfont\bbfam=\ninebb          \def\Bbb{\fam\bbfam\ninebb}%
\textfont\bffam=\ninebf          \scriptfont\bffam=\sixbf
\scriptscriptfont\bffam=\fivebf  \def\bf{\fam\bffam\ninebf}%
\abovedisplayskip=10pt plus 2pt minus 6pt \belowdisplayskip=%
\abovedisplayskip  \abovedisplayshortskip=0pt plus 2pt
\belowdisplayshortskip=5pt plus2pt minus 3pt  \smallskipamount=%
2pt plus 1pt minus 1pt  \medskipamount=4pt plus 2pt minus 2pt
\bigskipamount=10pt plus4pt minus 4pt  \setbox\strutbox=%
\hbox{\vrule height 7pt depth 2pt width 0pt}%
\normalbaselineskip=10pt \normalbaselines \rm}

\global\newcount\secno \global\secno=0 \global\newcount\meqno
\global\meqno=1 \global\newcount\appno \global\appno=0
\newwrite\eqmac \def\romappno{\ifcase\appno\or A\or B\or C\or D\or
E\or F\or G\or H\or I\or J\or K\or L\or M\or N\or O\or P\or Q\or
R\or S\or T\or U\or V\or W\or X\or Y\or Z\fi}
\def\eqn#1{ \ifnum\secno>0 \eqno(\the\secno.\the\meqno)
\xdef#1{\the\secno.\the\meqno} \else\ifnum\appno>0
\eqno({\rm\romappno}.\the\meqno)\xdef#1{{\rm\romappno}.\the\meqno}
\else \eqno(\the\meqno)\xdef#1{\the\meqno} \fi \fi
\global\advance\meqno by1 }

\global\newcount\refno \global\refno=1 \newwrite\reffile
\newwrite\refmac \newlinechar=`\^^J \def\ref#1#2%
{\the\refno\nref#1{#2}} \def\nref#1#2{\xdef#1{\the\refno}
\ifnum\refno=1\immediate\openout\reffile=refs.tmp\fi
\immediate\write\reffile{\noexpand\item{[\noexpand#1]\ }#2\noexpand%
\nobreak.} \immediate\write\refmac{\def\noexpand#1{\the\refno}}
\global\advance\refno by1} \def\semi{;\hfil\noexpand\break ^^J}
\def\nl{\hfil\noexpand\break ^^J} \def\refn#1#2{\nref#1{#2}}

\def\mplA#1#2#3{{\it Mod.\ Phys.\ Lett.}\ {\bf A{#1}} ({#2}) #3}

\def\pl#1#2#3{{\it Phys.\ Lett.}\ {\bf B{#1}} ({#2}) #3}
\def\plA#1#2#3{{\it Phys.\ Lett.}\ {\bf {#1}A} ({#2}) #3}

\newif\iftitlepage \titlepagetrue \newtoks\titlepagefoot
\titlepagefoot={\hfil} \newtoks\otherpagesfoot \otherpagesfoot=%
{\hfil\tenrm\folio\hfil} \footline={\iftitlepage\the\titlepagefoot%
\global\titlepagefalse \else\the\otherpagesfoot\fi}

\def\abstract#1{{\parindent=28pt\narrower\noindent\ninepoint\openup
2pt #1\par}}

\newcount\notenumber\notenumber=1 \def\note#1
{\unskip\footnote{$^{\the\notenumber}$} {\eightpoint\openup 1pt #1}
\global\advance\notenumber by 1}

\def\today{\ifcase\month\or January\or February\or March\or
April\or May\or June\or July\or August\or September\or October\or
November\or December\fi \space\number\day, \number\year}

\def\pagewidth#1{\hsize= #1}  \def\pageheight#1{\vsize= #1}
\def\hcorrection#1{\advance\hoffset by #1}
\def\vcorrection#1{\advance\voffset by #1}

\font\extra=cmss10 scaled \magstep0  \setbox1 = \hbox{{{\extra R}}}
\setbox2 = \hbox{{{\extra I}}}       \setbox3 = \hbox{{{\extra C}}}
\setbox4 = \hbox{{{\extra Z}}}       \setbox5 = \hbox{{{\extra N}}}





\def\frac#1#2{{#1\over#2}}

\def\pmb#1{\setbox0=\hbox{$#1$} \kern-.025em\copy0\kern-\wd0
    \kern.05em\copy0\kern-\wd0 \kern-.025em\raise.0433em\box0 }

\def\ve{\vfill\eject}

\def\R{{\Real}\!}

\def\({\left(}
\def\){\right)}


\def\R{{\rm I \hskip-0.47ex R}}


{

\refn\SchulmanRev
{L.S. Schulman, in \lq\lq Functional Integration
and its Applications\rq\rq, A.M. Arthurs, ed., 
Clarendon Press, Oxford, 1975}

\refn\Schulman
{L.S. Schulman, 
\lq\lq Techniques and Applications of 
Path Integration\rq\rq, John Wiley and Sons, 1981}

\refn\Cheng
{B.K. Cheng, \plA{101}{1984}{464}}


\refn\OursThree
{K. Horie, H. Miyazaki and I. Tsutsui, 
{\it Ann.\ Phys.} {\bf 279} (2000) 104}

\refn\OursTwo
{K. Horie, H. Miyazaki and I. Tsutsui, 
{\it Phys.\ Lett.} {\bf A253} (1999) 259}

\refn\Calogero
{F. Calogero, {\it J.\ Math.\ Phys.} {\bf 10} (1969) 2191, 
2197; {\bf 12} (1971) 419}

\refn\ChengChan
{B.K. Cheng and F.T. Chan, 
{\it J.\ Phys.} {\bf A20} (1987) 3771}

\refn\RS
{M. Reed and B. Simon, 
\lq\lq Methods of Modern
Mathematical Physics II, Fourier analysis, self-adjointness\rq\rq, 
Academic Press, New York, 1975}

\refn\AG
{N.I. Akhiezer and I.M. Glazman,
\lq\lq Theory of Linear Operators in Hilbert Space\rq\rq,
Vol.II,
Pitman Advanced Publishing Program, Boston, 1981}

\refn\AGHH
{S. Albeverio, F. Gesztesy, R. H{\o}egh-Krohn and H. Holden,
\lq\lq Solvable Models in Quantum Mechanics\rq\rq,
Springer, New York, 1988}

\refn\CFT
{T. Cheon, T. F\"{u}l\"{o}p and I. Tsutsui,
{\it Ann.\ Phys.} {\bf 294} (2001) 1}

\refn\Lathouwers
{L. Lathouwers, {\it J.\ Math.\ Phys.} {\bf 16} (1975) 1393}

\refn\Rellich
{F. Rellich, {\it Math.\ Ann.} {\bf 122} (1951) 343}

\refn\Krall
{A.M. Krall, {\it J.\ Differential Equations} {\bf 45} (1982) 128}


\refn\TsutsuiCheonFulop
{I. Tsutsui, T. Cheon and T. F\"{u}l\"{o}p, 
{\it Connection Conditions and the Spectral Family under Singular
Potentials}, in preparation}

\refn\Moebius
{I. Tsutsui, T. F\"{u}l\"{o}p and T. Cheon,
{\it J.\ Math.\ Phys.} {\bf 42} (2001) 5687}

\refn\PeakInomata
{D. Peak and A. Inomata, {\it J.\ Math.\ Phys.} {\bf 10} 
(1969) 1422}

\refn\WoottersZurek
{W.K. Wootters and W.H. Zurek,
{\it Nature}  {\bf 299} (1982) 802}

\refn\AY
{Y. Akaishi and T. Yamazaki, 
{\it Nuclear $\bar K$ Bound States in Light Nuclei},
KEK Preprint 2001-48, {\it Phys.\ Rev.} {\bf C},
to appear}

\refn\Gov
{T. Govindarajan, V. Suneeta and S. Vaidya, 
{\it Nucl.\ Phys.} {\bf B583} (2000) 291}

\refn\Birm
{D. Birmingham, K. Gupta and S. Sen, 
\pl{505}{2001}{191}}

\refn\Mignemi
{S. Mignemi, 
\mplA{16}{2001}{1997}}

\refn\Erdelyi
{A. Erdelyi, ed., \lq\lq Higher Transcendental Functions\rq\rq{} 
Vol.II, McGraw-Hill, New York, 1953}

}



\pageheight{23cm}
\pagewidth{14.8cm}
\hcorrection{0mm}
\magnification= \magstep1
\def\bsk{%
\baselineskip= 16.8pt plus 1pt minus 1pt}
\parskip=5pt plus 1pt minus 1pt
\tolerance 6000


\null

{
\leftskip=100mm
\hfill\break
KEK Preprint 2001-164
\phantom{KEK Preprint 2001-164}
\hfill\break
\par
}

\vskip 30pt

{\baselineskip=18pt

\centerline{\bigbold
Quantum Tunneling and Caustics}
\centerline{\bigbold
under Inverse Square Potential}

\vskip 30pt

\centerline{\smc
Hitoshi Miyazaki\footnote{${}^*$}
{\eightpoint email:\quad miyazaki@post.kek.jp}
\quad
{\rm and}
\quad
Izumi Tsutsui\footnote{${}^\dagger$}
{\eightpoint email:\quad izumi.tsutsui@kek.jp}
}

\vskip 20pt

{
\baselineskip=13pt
\centerline{\it
Institute of Particle and Nuclear Studies}
\centerline{\it
High Energy Accelerator Research Organization (KEK)}
\centerline{\it
Tsukuba 305-0801}
\centerline{\it Japan}
}

\vskip 80pt

\abstract{%
{\bf Abstract.}\quad
We examine the quantization of a harmonic oscillator with inverse square potential 
$V(x) = ({m\omega^2}/{2}) \,{x^2} + g/{x^2}$
on the line $ -\infty < x < \infty$.  We find that, for 
$0 < g < 3\hbar^2/(8m)$, the system admits a $U(2)$ family of
inequivalent quantizations allowing for 
quantum tunneling through the infinite 
potential barrier at $x=0$.  These are a generalization of the 
conventional quantization applied to the
Calogero model in which no quantum tunneling is allowed.  
The tunneling renders the classical caustics which arise under the potential 
anomalous at the quantum level, leading to the possibility of copying the
profile of an arbitrary state from one side
$x > 0$, say, to the other $x < 0$.
}

\vskip 20pt
{\baselineskip=12pt
{\ninepoint
\indent{PACS codes: 03.65.-w, 73.63.-b, 02.30.-f \hfill\break}
}
}

%
%
%
%
%

\ve


\pageheight{23cm}
\pagewidth{15.7cm}
\hcorrection{-1mm}
\magnification= \magstep1
\def\bsk{%
\baselineskip= 15pt plus 1pt minus 1pt}
\parskip=5pt plus 1pt minus 1pt
\tolerance 8000
\bsk


\bigskip
\noindent{\bf 1. Introduction}
\medskip

In some dynamical systems there occurs a peculiar phenomenon that
classical trajectories 
focus on one single point 
after a lapse
of a certain time irrespective of their initial
conditions.  This occurs typically in a harmonic oscillator, 
where the oscillator returns to the initial 
position periodically
whatever its initial velocity is.  
This phenomenon underlies the classical caustics of geometrical
optics, whose quantum version has also been
studied earlier
[\SchulmanRev, \Schulman]. 
Since the phenomenon is genuinely classical, 
one is tempted to ask if any substantial change occurs in
the caustics at the quantum level.  A path-integral analysis [\Cheng,
\OursThree] indicates, however, that for quadratic systems the
focusing phenomenon remains essentially unchanged after quantization --- 
it arises as a recurrence of the initial profile of probability
distributions, accompanied by certain quantum effects [\OursTwo].

The caustics phenomenon remains to be seen, at least
classically, even when an inverse square potential
is added to the harmonic oscillator.  Indeed, the Hamiltonian,
$$
H(p, x) =\frac{1}{2m} {p^2} + \frac{m {\omega^2}}{2} {x^2} +
g\frac{1}{x^2},
\eqn\Hamiltonian
$$
on the line $ -\infty < x < \infty$ admits
classical solutions which exhibit
periodicity for positive
strength $g > 0$, implying that caustics still occur despite that the
system is now non-quadratic.
This system is in fact the quantum solvable model considered by Calogero 
[\Calogero] and, like in quadratic systems,
has been argued to exhibit corresponding caustics at the
quantum level [\ChengChan].  The
argument is based on the conventional quantization of the system (\Hamiltonian) 
used for the Calogero model which assumes
no probability flow to pass the singular (infinite)
barrier at $x = 0$. 
Because of this prohibition of quantum tunneling, 
the system does
not reduce to the harmonic oscillator for $g
\to 0$ as one na\"\i vely expects.

On the other hand, it has been known in 
the mathematical literature that 
systems with singularity such as 
the one mentioned above may
have inequivalent quantizations due to 
the arbitrariness of the boundary (or
connection) condition at the singularity (see, {\it e.g.}, 
[\RS, \AG] for
systems on the half line, and [\AGHH, \CFT] 
for those with point interaction). 
In this paper we examine the quantization 
of the system (\Hamiltonian) from
this viewpoint and point out that, for $g$ in the range, 
$$
0 < g < \frac{3\hbar^2}{8m},
\eqn\gregion
$$ 
the system admits a $U(2)$ family of inequivalent
quantizations.  In contrast to the conventionl one, most of these quantizations permit
quantum tunneling through the infinite barrier at $x = 0$.  In particular, we find that there is 
a distinguished quantization possessing 
a smooth limit to the harmonic oscillator for $g
\to 0$.   Remarkably, in this quantization the classical picture of
the caustics acquires a drastic change in the quantum
regime: the
focusing occurs in two points, rather than one.   
We show that this anomalous phenomenon of quantum
caustics may be used to   
copy the profile of an arbitrary state from one side $x > 0$,
say, to the other $x < 0$.

\bigskip
\noindent{\bf 2. Classical caustics and quantum states}
\medskip

Before delving into the discussion of quantization of
the system (\Hamiltonian), let us quickly recall how the
phenomenon of the  classical caustics can be observed.  Let
$H(p, x) = E$ be the energy of the solution we are looking for. 
With $p = m\dot x$ the constant energy equation
can readily be integrated to give the classical
solution,
$$
x(t)=\pm\left\{\sqrt{\frac{{E^2}-2gm{\omega^2}} {(m \omega^2) 
^2 }}
\sin(2\omega(t+{t_0}))+\frac{E} {m \omega^2}\right\}^{1/2}.
\eqn\XCl
$$
The integration constant $t_0$ and the energy $E$ are 
altered according to
the initial position $x(t_i)$ and 
velocity $\dot x(t_i)$ chosen arbitrarily, but for
any choice the particle returns to
the original position $x(t_i + T) = x(t_i) $ for $T =
k\pi/\omega$ with integer
$k$.  These are the classical caustics appearing in the system
(\Hamiltonian).

The quantum system corresponding to (\Hamiltonian), too, admits exact
solutions of eigenstates for the 
Schr{\"o}dinger 
equation.  Although
the procedure to obtain the solutions has been given in various
references (see, {\it e.g.}, [\Calogero, \Lathouwers]), we
shall present here a fuller treatment paying a special attention to the
boundary condition at $x = 0$.  To proceed,
we remove the singular point $x = 0$ from the system to define our
Hilbert space as
${\cal H} = L^2(\R\!\setminus\!\{0\})$.  The boundary condition is
then considered at the limiting points $x \to \pm 0$.
For the Hamiltonian operator $\widehat H = H(-i\hbar d/dx, x)$ 
the Schr{\"o}dinger equation for energy eigenstates reads
$$
\widehat H {\psi_n}(x) = \left(-\frac{\hbar^2}{2m}\frac{d^2}{d
x^2}+\frac{m
\omega^2}{2}{x^2} +g\frac{\>1\>}{x^2}\right)
\psi_n  (x)={E_n}{\psi_n}(x).
\eqn\Schroedinger
$$ 
For the moment we only consider the positive half line
$\R_+ = \{x > 0\}$, but the negative half
$\R_- = \{x < 0\}$ can be handled analogously using the solutions on $\R_+$.  
If we set
$$
\psi_n  (x)={y^{a + 1/2}} e^{-{y^2}/2}\,
{f_n}({y^2}), \qquad
y =\sqrt{\frac{m\omega}\hbar} \, x,
\eqn\Ansatz
$$
and choose 
$$
a=
{1\over 2}\sqrt{1+\frac{8mg}{\hbar^2}},
\eqn\svalue
$$
then the Schr{\"o}dinger equation
(\Schroedinger) becomes
$$
z \frac{d^2f_n}{dz^2}(z)
+\left(a + 1 -z\right)\frac{d \,f_n}{dz}(z)
-\frac{1}{2}\left(a + 1 -\lambda_n\right)
f_n(z)=0, \qquad
\lambda_n = \frac{E_n}{\hbar\omega},
\eqn\chgeqation
$$
under the variable $z = y^2$.  This is 
the confluent
hypergeometric differential equation 
$z f''(z)+(\gamma-z) f'(z)-\alpha
f(z)=0$, whose
two independent solutions 
are, for $\gamma\neq \rm{integer}$, given by
$
f(z)=F(\alpha,\gamma;z)
$
and
$
z^{1-\gamma} F(\alpha-\gamma+1,2-\gamma;z)
$
with $F(\alpha,\gamma;z)$ being the confluent hypergeometric
function.  
Thus 
the two independent solutions for the Schr{\"o}dinger equation 
(\Schroedinger) are
$$
\eqalign{
\phi^{(1)}_n(x) &:={y^{c_1 - 1/2}}e^{-{y^2}/{2}}
F\left(\frac{{c_1}-\lambda_n}2,{c_1};{y^2}\right), \qquad c_1 = 1 + a,\cr
{\phi^{(2)}_n}(x) &:={y^{c_2 - 1/2}}e^{-{y^2}/{2}}
F\left(\frac{{c_2}-\lambda_n}2,{c_2};{y^2}\right), \qquad c_2 = 1 - a.
}
\eqn\Solution
$$
The general solution $\psi_n (x)$ of (\Schroedinger) is given by a linear
combination of these two, but since the combination may differ on the two sides
$\R_+$ and
$\R_-$, we have
$$
\psi_n (x)
=[{N_{\rm R}^{(1)}}\phi^{(1)}_n(|x|)+{N_{\rm R}^{(2)}}\phi^{(2)}_n(|x|)]\Theta(x)
+[{N_{\rm L}^{(1)}}\phi^{(1)}_n(|x|)+{N_{\rm
L}^{(2)}}\phi^{(2)}_n(|x|)]\Theta(-x),
\eqn\solutions
$$
where $N_{\rm R}^{(s)}$ and $N_{\rm L}^{(s)}$ 
are arbitrary
constants and
$\Theta(x)$ is the Heaviside step function.

At this point let us examine the normalizability (square integrability) of the
solutions (\Solution).  First, since $c_1 > 3/2$ and $c_2 < 1/2$, we
observe that as $x
\to 0$ the solution
$\phi^{(1)}_n$ approaches zero while 
$\phi^{(2)}_n$ diverges.  From 
$\int^\epsilon_0 dx |\phi^{(2)}_n (x) |^2 \simeq \epsilon^{2c_2}$ for a small
$\epsilon$, we realize that $\phi^{(2)}_n$ can be normalizable for $c_2 > 0$. 
This is the case if the coupling constant $g$ satisfies
(\gregion), and we confine ourselves to this case hereafter.  (For the
normalizability $g$ may be non-positive, but for our consideration of
quantum tunneling and caustics we assume $g > 0$.)  Note
that (\gregion) implies 
$3/2<{c_1}<2$ and $0<{c_2}<1/2$, and this allows us to disregard the
case
$\gamma = {\rm integer}$ in considering the solution of (\chgeqation).
Once the two independent solutions are admitted from the
behaviour near $x = 0$, then the normalizability is ensured if the
solution vanishes sufficiently fast at the infinity $x = \pm \infty$. 
From the asymptotic behaviour of the confluent hypergeometric function, 
$$
F(\alpha,\gamma;z)\sim\frac{\Gamma(\gamma)}
{\Gamma({\alpha})}{e^z}{z^{\alpha-\gamma}},\qquad
\hbox{as}\quad  |z| \rightarrow \infty,
\eqn\no
$$
we find that the normalizability of the solutions (\Solution) requires
$$
\frac{N_{\rm R}^{(1)}}{N_{\rm R}^{(2)}}=\frac{N_{\rm L}^{(1)}}{N_{\rm L}^{(2)}}=
-\frac{\Gamma\left(({c_1}-\lambda_n)/2\right)}
{\Gamma\left(({c_2}-\lambda_n)/2\right)}
\frac{\Gamma(c_2 )}{\Gamma(c_1)}.
\eqn\Ratio
$$

Another condition to be imposed on
the solutions is the boundary condition at the singular point
$x = 0$.  This is needed to 
ensure the continuity of the probability current 
at $x = 0$, which is equivalent to ensuring that the
Hamiltonian $\widehat H$ be self-adjoint.  
It is known [\RS, \AG] that, in the presence of singularity, there
can exist (at most) a $U(2)$ family of self-adjoint
Hamiltonians specified by corresponding 
boundary conditions.   
By means of the Wronskian 
$W[\psi,\varphi](x) = (\psi (d \varphi/dx)
-(d\psi/dx)\varphi)(x)$, 
which is finite even if the wavefunctions $\psi(x)$, $\varphi(x)$
may be divergent at the singularity, the boundary conditions
are presented as follows [\Rellich, \Krall]  (see
[\TsutsuiCheonFulop, \Moebius] for the conditions on the line).   Let
$\varphi_{1}$, 
$\varphi_{2}$ be two independent, real zero modes,
$$
\widehat H\varphi_{1}(x) = \widehat H\varphi_{2}(x) = 0,
\qquad W[\varphi_{1},\varphi_{2}](x) = 1.
\eqn\zeromode
$$
Given a state $\psi$ which is normalizable,   
we introduce the complex column
vectors, 
$$
\Psi=\pmatrix{
W[\psi,\varphi_{1}]_{+0}\cr
W[\psi,\varphi_{1}]_{-0}
},
\qquad
\Psi'=\pmatrix{
W[\psi,\varphi_{2}]_{+0}\cr
- W[\psi,\varphi_{2}]_{-0}
},
\eqn\boundvect
$$
defined from the boundary values
$W[\psi,\varphi]_{\pm 0} := \lim_{x \to \pm 0}W[\psi,\varphi](x)$. 
The boundary condition for $\psi \in {\cal H}$ is then
given by
$$
(U-I)\Psi+i {L_0}(U+I)\Psi' =0,
\eqn\BoundaryCondition
$$
where $U$ is a $U(2)$ matrix, 
$I$ is the identity matrix, and 
$L_0$ is a constant with 
dimension of length.  This way a self-adjoint Hamiltonian is
specified uniquely by the matrix
$U$, which may hence be called the \lq characteristic matrix\rq.

In our case, we label $n = n_0$ for which $\lambda_{n_0} = 0$ in
(\Solution) and set
$$
\eqalign{
\varphi_{1}(x) &:= \sqrt{\frac{\hbar}{m\omega}}\,
\phi_{n_0}^{(1)}(\vert x\vert)\left[\Theta(x) -
\Theta(-x)\right],\cr
\varphi_{2}(x) &:= 
\frac{1}{c_2  -  c_1}\,
\phi_{n_0}^{(2)}(\vert x\vert),
}
\eqn\no
$$
so that (\zeromode) is fulfilled.  Since
$F(\alpha,\gamma;z)= 1 + {\cal O}(z)$ as $z \to 0$,
the boundary vectors (\boundvect) for the solution
$\psi_n$ in (\solutions) turn out to be
$$
\Psi= ({c_1}-{c_2})
\pmatrix{
N_{\rm R}^{(2)}\cr
N_{\rm L}^{(2)}
},
\qquad
\Psi'=\sqrt{\frac{m\omega}\hbar} \, 
\pmatrix{
N_{\rm R}^{(1)}\cr
N_{\rm L}^{(1)}
}.
\eqn\Vectors
$$
The relations 
(\Ratio) and (\Vectors) then imply that the vector 
$\Psi'$ is
proportional  to $\Psi$, and hence 
there exists a constant $\xi$ such that
$\Psi'=\xi \Psi$. 
Thus the boundary condition (\BoundaryCondition) is now
$$
\left[(U-I)+i{L_0}\xi(U+I)\right]\Psi=0,
\eqn\boundarycon
$$
and, in order to obtain a 
non-trivial vector $\Psi$, we need
$$
\det
\left|
U-I+i{L_0}\xi(U+I)
\right|
=\det\left|
D-I+i{L_0}\xi(D+I)
\right|
=0,
\eqn\BoundaryConditionTwo
$$
where we have decomposed $U \in U(2)$ as 
$U=V^{-1}DV$ using some $SU(2)$ matrix $V$ and a diagonal
matrix $D$.  In terms of the parameterization, 
$$
D=
\pmatrix{
e^{i{\theta_{+}}} & 0\cr
0 & e^{i{\theta_{-}}}
},
\eqn\no
$$
with
$\theta_\pm \in
[0, 2\pi)$,
we find that (\BoundaryConditionTwo) is satisfied if
$$
\xi =-\frac{1}{L_+} \quad\hbox{or}\quad -\frac{1}{L_-}, \qquad
\qquad
{L_\pm}={L_0}\cot
\left(
\frac{{\theta_\pm}}{2}
\right).
\eqn\no
$$
Substituting this back to (\Ratio), we obtain 
$$
\frac{\Gamma\left(({c_1}-\lambda_n)/2\right)}
{\Gamma\left(({c_2}-\lambda_n)/2\right)}
\frac{\Gamma(c_2 )}{\Gamma(c_1)}
=\sqrt{\frac\hbar{m\omega}}
\frac{{c_1}-{c_2}}{L_+}
\quad  \hbox{or} \quad 
\sqrt{\frac\hbar{m\omega}}
\frac{{c_1}-{c_2}}{L_-},
\eqn\EnergyDeterminedOne
$$
from which we determine the energy spectrum $\{E_n = 
\lambda_n \hbar \omega\}$ of our system.  
The ratios $N_{\rm R}^{(1)}/N_{\rm R}^{(2)}$ and $N_{\rm L}^{(1)}/N_{\rm L}^{(2)}$ are determined
once either $L_+$ or $L_-$ is chosen.   Our result shows that the
system permits two distinct series of eigenstates generically, one
specified by
$L_+$ and the other by
$L_-$, and this illustrates the fact that any one dimensional system
which admits a $U(2)$ family of self-adjoint Hamiltonians possesses a
spectral family parametrized by two angle parameters
[\TsutsuiCheonFulop, \Moebius].

We shall mention a few cases where the spectrum $\{E_n\}$ can be obtained
explicitly.
First, if $(\theta_{+},\theta_{-})=(0,0)$, then $1/L_\pm = 0$ and 
hence (\EnergyDeterminedOne) is fulfilled by those $\lambda_n$ for which
the Gamma function in the denominator has poles.  This leads to 
$E_n = (2n+{c_2})\hbar\omega$ and the eigenstate given by 
$\phi_n^{(2)}(\vert x\vert)$ either on $\R_+$ or 
$\R_-$  (hence each level is doubly degenerated).  Similarly, if
$(\theta_{+},\theta_{-})=(\pi,\pi)$, then $L_\pm = 0$ and
one obtains 
$E_n = (2n+{c_1})\hbar\omega$ and the eigenstate 
$\phi_n^{(1)}(\vert x\vert)$ which is also doubly degenerated.  This is the
case (which amounts to the choice $U = -I$) that has been considered
conventionally in the treatment of the system (\Schroedinger) since the early
days of Calogero [\Calogero].

On the other hand, if
$(\theta_{+},\theta_{-})=(0,\pi)$, then $1/L_+ = 0 = {L_-}$, which 
means that there occurs two series of eigenstates, one 
with ${N_{\rm R}^{(2)}}={N_{\rm L}^{(2)}}=0$ and the other with
${N_{\rm R}^{(1)}}={N_{\rm L}^{(1)}}=0$, 
whose eigenvalues are 
$$
E^{(1)}_n =(2n+{c_1})\hbar\omega, \qquad
E^{(2)}_n =(2n+{c_2})\hbar\omega,
\eqn\evalues
$$
respectively.  In particular, in the
limit $g \to 0$ we have $c_1 \to 3/2$ and $c_2 \to 1/2$, which shows that
our system recovers the spectrum of
a harmonic oscillator.  A complete reduction to the harmonic oscillator
system is realized by choosing $U = \sigma_1$ (where $\{\sigma_i\}$ are
Pauli matrices), which is obtained by setting $V = e^{i\pi\sigma_2/4}$
as well as
$(\theta_{+},\theta_{-})=(0,\pi)$.  For this choice, the boundary condition
(\boundarycon) requires
${N_{\rm R}^{(1)}}=-{N_{\rm L}^{(1)}}$, ${N_{\rm R}^{(2)}}={N_{\rm
L}^{(2)}}$ and hence the two series of eigenstates in (\solutions) are found
to be
$$
\eqalign{
\psi_n^{(1)}(x)
&:=
N^{(1)}\, \phi^{(1)}_n(\vert x\vert) 
\left[\Theta(x) - \Theta(-x)\right], \cr 
\psi_n^{(2)}(x)
&:=
N^{(2)}\, \phi^{(2)}_n(\vert x\vert) ,
}
\eqn\SolutionTwo
$$ 
for $n = 0$, 1, $2\ldots\,$, where
$N^{(s)} = [\sqrt{mw/\hbar} \,\Gamma(n + c_s)/
\{(\Gamma(c_s))^2 n!\}]^{1/2}$ for $s = 1$, 2 are normalization constants
determined so that $\int_{-\infty}^\infty dx\, 
\vert \psi_n^{(s)}(x)
\vert^2 = 1$. 
The eigenfunctions (\SolutionTwo) reduce exactly to
those of the harmonic oscillator in the limit 
$g \rightarrow 0$, that is, $\psi_n^{(1)}$ reduces to
$e^{-y^2/2}H_{2n+1}(y)$ 
and $\psi_n^{(2)}$ to $e^{-y^2/2}H_{2n}(y)$ where $H_n$ is the
Hermite polynomial of degree $n$. This in turn implies that, for other
$U$, the system does not lead to a harmonic oscillator
in the limit, which suggests that 
our system with finite
$g$ may be regarded, effectively, as a system that possesses a
singular point interaction at
$x = 0$ which is hidden in the singularity of the
potential.   Under regular
potentials, point interactions are known to admit a $U(2)$
family of boundary conditions at the singularity, in which
$U = \sigma_1$ provides the boundary condition for 
the \lq free point\rq{}, namely, no interaction there [\CFT].  The
fact that the smooth limit $g \to 0$ to the harmonic
oscillator is gained at $U = \sigma_1$ suggests that the above effective
picture for the $U(2)$ family works also for singular potentials.
We also mention that the case $U = \sigma_1$ corresponds to the
quantization discussed in ref.[\Lathouwers] which pointed out
that the conventional quantization $U = -I$ cannot be a
perturbed harmonic oscillator because of the too severe
physical conditions it presupposes.

\bigskip
\noindent{\bf 3. Quantum caustics and its anomaly}
\medskip

Now that we have unconventional but perfectly admissible eigenstates arising
under the boundary conditions specified by $U$, we next examine how the
caustics appear at the quantum level.  
Before this, however, let us consider the possibility of quantum tunneling
though the barrier of the potential at $x = 0$.   
In order to make our
discussions clear and simple,  we consider only the case 
$U = \sigma_1$ where the eigenstates are given by 
(\SolutionTwo).   
To investigate whether or not tunneling phenomena occurs, we simply
evaluate the  probability current $j(+0) (=
j(-0))$ for a given
arbitrary state
$\psi$. Since (\SolutionTwo) gives our complete basis, we expand it as
$\psi(x) = \sum_n(c_n^{(1)}\psi_n^{(1)}(x) +
c_n^{(2)}\psi_n^{(2)}(x))$.  
Then we find
$$
j(+0) := \frac{\hbar}{2im}W[\psi^*,\psi]_{+0} 
=  \frac{ia\hbar}{m}\sum_{n, l} 
\left\{ (c_n^{(1)})^*c_l^{(2)} - (c_n^{(2)})^*c_l^{(1)}\right\},
\eqn\prcurrrent
$$
which shows that, since $s \ne 0$ for $g > 0$, the
probability current does flow through the barrier
$x = 0$.  Note that $j(+0) \ne 0$ is realized for states $\psi$
consisting of both type of eigenstates $\psi_n^{(1)}$ and
$\psi_l^{(2)}$, and this is made possible only for $g$ satisfying
(\gregion) and further for (generic) $U$, such as the one
$U = \sigma_1$ we are considering, under which the two type of
eigenstates appear.  If $g \ge 3 \hbar^2  /8$, or else if $U$ is
diagonal $U = D$ like the conventional choice $U = -I$, we always
have $j(+0) = 0$, disconnecting the right and left half lines, $x >0$
and
$x < 0$, physically.

Once the quantum tunneling is allowed, then the classical picture
of caustics, which occur in the half lines independently, is no
longer viable, and one is curious what in fact will happen quantum
mechanically.  To investigate this, we calculate the transition
amplitude, the Feynman kernel $K({x_f},{t_f};{x_i},{t_i})$, from
the initial state of the particle staying at $x = x_i$ at $t = t_i$
to the final state staying at $x = x_f$ at $t = t_f$.
In our case
(\SolutionTwo), a straightforward computation 
(see Appendix) yields that for 
$T := t_f - t_i \neq k\pi/\omega $
with
$k = 0, 1, 2, \ldots\,$,
$$
\eqalign{
&K({x_f},{t_f};{x_i},{t_i})
=\frac{m\omega}{2i\hbar\sin(\omega T)}
(|{x_f}{x_i}|)^{1/2}
\exp\left(
\frac{i}{2}\frac{m\omega}{\hbar}
\frac{\cos(\omega T)}{\sin(\omega T)} 
({x^2_f}+{x^2_i})
\right)\cr
&\qquad\qquad\times 
\biggl[
\Theta(x_fx_i)
\left\{I_{a}
\left(
\frac{m\omega}{i\hbar}\frac{|{x_f}{x_i}|}{\sin(\omega T)}
\right)
+I_{-a}
\left(
\frac{m\omega}{i\hbar}\frac{|{x_f}{x_i}|}{\sin(\omega T)}
\right)
\right\}\cr
&\quad\qquad\qquad +
\Theta(-x_fx_i)
\left\{-I_{a}
\left(
\frac{m\omega}{i\hbar}\frac{|{x_f}{x_i}|}{\sin(\omega T)}
\right)
+I_{-a}
\left(
\frac{m\omega}{i\hbar}\frac{|{x_f}{x_i}|}{\sin(\omega T)}
\right)
\right\} 
\biggr],
}
\eqn\FeynmanKernelOne
$$
where $I_\nu(z)$ 
is the modified Bessel function and $a$ is related to
$g$ by (\svalue).  The last two terms with the factor
$\Theta(-x_fx_i)$
represent the transition allowed by the quantum tunneling.
One can readily check 
that the Feynman kernel (\FeynmanKernelOne) reduces to that of a 
harmonic oscillator in the limit $g\rightarrow 0$. 

On the other hands, for
$T=k\pi/\omega $, we find
$$
K({x_f},{t_f};{x_i},{t_i}) 
=
(-1)^k \cos(ak\pi)\delta({x_f}-{x_i})
+ i(-1)^k \sin(ak\pi)\delta({x_f}+{x_i}).
\eqn\FeynmanKernelTwo
$$
The term containing $\delta({x_f}-{x_i})$ represents the quantum counterpart
of the classical caustics, whereas the term containing 
$\delta({x_f}+{x_i})$ represents extra caustics that
arise only at the quantum level through the tunneling effect.
We emphasize that the appearance of the anomalous quantum caustics is
crucial to achieve the smooth reduction to the harmonic oscillator,
since 
$g\rightarrow 0$ implies $a \to 1/2$ and hence the two terms contribute to the
caustics of the harmonic oscillator alternately.  

In passing we note that
the other limit $a \to 1$ is also smooth, because then the kernel, 
(\FeynmanKernelOne) or (\FeynmanKernelTwo), becomes the usual one 
[\PeakInomata] (since, for 
(\FeynmanKernelOne) the last two terms with 
$\Theta(-x_fx_i)$ cancel with each other, whereas for 
(\FeynmanKernelTwo) we only get
$\delta({x_f}-{x_i})$)
obtained
under the conventional quantization.  This is due to the fact that, in our
treatment,
the second
solution $\psi_n^{(2)}(x)$ in (\SolutionTwo) ceases to exist formally 
as $a \to 1$ because of the normalization factor $N^{(2)}$. 

%
%
%
\topinsert
\vskip 0.5cm
\let\picnaturalsize=N
\def\picsize{10cm}
\def\picfilename{f1.epsf}
\input epsf
\ifx\picnaturalsize N\epsfxsize \picsize\fi
\hskip 1.2cm\epsfbox{\picfilename}
\vskip 0.5cm
\abstract{%
{\bf Figure 1.}\quad
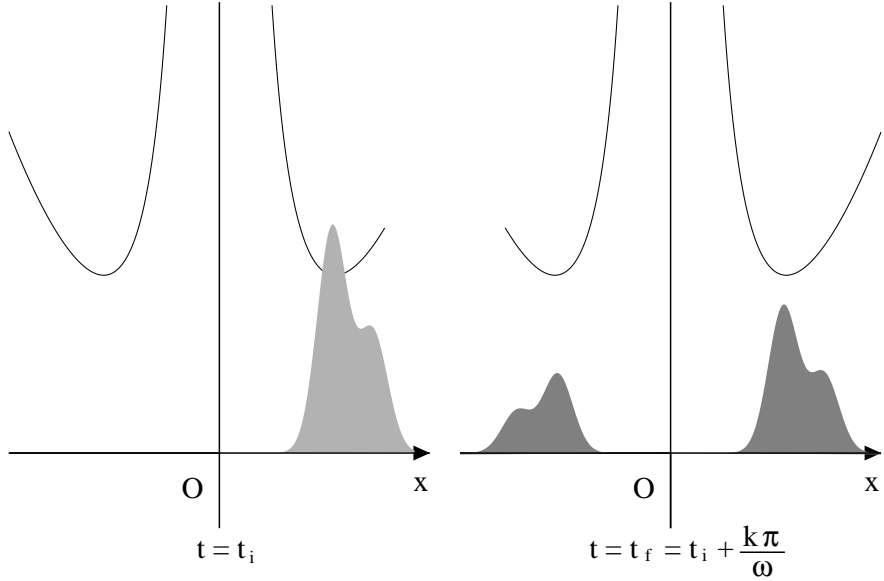
Process of quantum copy through the caustics anomaly.  At every period $T =
k\pi/\omega$, a mirror image of the original profile on $x > 0$ 
emerges on the
other side $x < 0$.  The relative 
size of the mirror image depends on $a$ and
$k$.}
\endinsert

In order to see the physical consequence of the caustics anomaly,
let us consider an initial state $\psi(x, t_i)$ 
whose density $\rho_i(x) = \vert \psi(x, t_i) \vert^2$
has a support only on $\R_+$.  The state
evolves according to the rule set by the kernel (\FeynmanKernelOne), and
hence the profile will broaden and enter in $\R_-$ at some later
time.   The salient feature of the usual quantum caustics observed for
quadratic systems is that, at
$t_f = t_i + T$ with $T = (\hbox{period of caustics}) \times
\hbox{integer}$, the initial profile is reproduced completely.  In our
system, however, this is no longer true because for
$T = k\pi/\omega$ 
we have the final state
$\psi(x, t_f) 
= \int dx'\, 
K({x},{t_f};{x'},{t_i}) \, \psi(x', t_i)
$
with the density,
$$
\rho_f(x) 
= \vert \psi(x, t_f) \vert^2
= \cos^2(ak\pi)\,\rho_i(x) + \sin^2(ak\pi)\, \rho_i(-x).
\eqn\StateAQ
$$
This shows that, at any later periods, the profile on $\R_+$ is copied as a mirror
image on 
$\R_-$ (see Figure
1).  In particular, when $a = 3/4$ ({\it i.e.}, $g = 5\hbar^2/(32m)$), the mirror image
becomes exactly of the same size as the original for odd $k$, while for even
$k$ the complete
profile is reproduced on $\R_-$ and $\R_+$ alternately.   
We note that this does not contradict the 
no-cloning theorem [\WoottersZurek] because the
two \lq state spaces\rq{} on the half lines, 
$L^2(\R_+)$ and $L^2(\R_-)$, do not comprise
the entire Hilbert space by their direct product,
${\cal H} = L^2(\R\setminus\{0\}) \ne L^2(\R_+)\otimes L^2(\R_-)$.  
In short,  rather than making a replica of an arbitrary state prohibited by the
no-cloning theorem, the above copying process duplicates a profile by the
mirror image.

Since the system discussed in this paper arises in various branches of
physics, we expect that our result will find several other
applications, and to conclude we just mention a few.  
First, if one is to confine a particle with more than one
channels among which the probability can flow like in certain
nuclear states [\AY] or nano-devices with spin channels, then our
quantizations may be adequate to apply.   
The second is the analysis of black holes, where our system (with and
without the harmonic term) describes a particle probing the 
near-horizon geometry [\Gov, \Birm, \Mignemi].  
Further, a straightforward extension of the
quantizations of 
the $n$-body Calogero
model (and its related solvable models) along the line outlined here would
also enlarge the scope of the application of the model on account of the
quantum tunneling now allowed.

\bigskip
\noindent
{\bf Acknowledgement:}
I.T.~is indebted to T. Cheon and R. Sasaki for useful comments.
This work has been supported in part by
the Grant-in-Aid for Scientific Research 
(Nos.~10640301 and 13135206) by
the Japanese
Ministry of
Education, Science, Sports and Culture.

\secno=0 \appno=1 \meqno=1

\bigskip
\noindent{\bf Appendix}
\medskip

In this Appendix we calculate the Feynman kernel
$K({x_f},{t_f};{x_i},{t_i})$ from the energy
eigenfunctions (\SolutionTwo).  Putting $T={t_f}-{t_i}$ it is given by
$$
K({x_f},{t_f};{x_i},{t_i})
= \langle x_f \vert e^{-i{\widehat H}T/\hbar} \vert
x_i\rangle
={S^{(1)}}+{S^{(2)}},
\eqn\inikernel
$$
with
$$
{S^{(s)}}=\sum^\infty_{n=0} {\psi^{(s)}_n} ({x_f})\,
e^{-\frac{i}{\hbar}E^{(s)}_n T}\,({\psi^{(s)}_n}(x_i))^*, \qquad
s = 1, \, 2.
\eqn\esone
$$
To evaluate ${S^{(1)}}$, we plug (\SolutionTwo) into (\esone)
using dimensionless variables $y_i = \sqrt{m\omega/\hbar}
\, x_i$, $y_f = \sqrt{m\omega/\hbar}
\, x_f$ to find
$$
\eqalign{
{S^{(1)}}
&= 
\sqrt{\frac{m\omega}{\hbar}}(|{y_f}{y_i}|)^{{c_1}-1/2}
\,e^{-\frac{1}{2}({y^2_f}+{y^2_i})} \,
\left[\Theta(y_fy_i) - \Theta(-y_fy_i)\right]
\cr
& 
\qquad\times
\sum^\infty_{n=0}e^{-i(2n+{c_1})\omega T}\frac{n!}{\Gamma({c_1}+n)}
{L^{({c_1}-1)}_n}({y^2_f}){L^{({c_1}-1)}_n}({y^2_i}),
}
\eqn\Splus
$$
where we have used the relation between the confluent hypergeometric
functions and the (associated) Laguerre
polynomials,
$$
F\left(-n,{\gamma};{z}\right)
=\frac{\Gamma({\gamma})\, n!}{\Gamma({\gamma}+n)}
{L^{({\gamma}-1)}_n}({z}).
\eqn\no
$$
Employing the standard trick $T \rightarrow T(1-i\epsilon)$ with an infinitesimal $\epsilon$
in (\Splus) to ensure
the convergence of the kernel, and
using the Hill-Hardy formula (see p.189, ref.[\Erdelyi]),
$$
\eqalign{
\sum^\infty_{n=0} & 
{w^n} \frac{n!}{\Gamma(\nu+n+1)}
{L^{(\nu)}_n}(u)
 {L^{(\nu)}_n}(v)\cr
&\qquad = \left(\frac{1}{1-w}\right)
\exp\left(-w\frac{u+v}{1-w}\right)(uvw)^{-\nu/2}
I_\nu
\left(2\frac{(uvw)^{1/2}}{1-w}\right),
}
\eqn\HillHardy
$$
valid for $|w| < 1$, where $I_\nu (z)$ denotes the first kind of
the modified Bessel function, 
we obtain
$$
\eqalign{
S^{(1)} 
&=  \lim_{\epsilon \to +0}
\sqrt{\frac{m\omega}{\hbar}}(|{y_f}{y_i}|)^{{c_1}-1/2}
\,e^{-\frac{1}{2}({y^2_f}+{y^2_i})} \,
\left[\Theta(y_fy_i) - \Theta(-y_fy_i)\right]\,e^{-i\omega T{c_1}}
\cr
&\qquad\qquad\times
\frac{e^{c_1\epsilon/2}}{1-e^{-i2\omega T-\epsilon}}
\exp\left(-e^{-i2\omega T-\epsilon}
\frac{{y^2_f}+{y^2_i}}{1-e^{-i2\omega T-\epsilon}}
\right)\cr
&\qquad\qquad\times\left({y^2_f}{y^2_i}e^{-i2\omega 
T-\epsilon}\right)^{-({c_1}-1)/2}
I_{{c_1}-1}\left(2\frac{({y^2_f}{y^2_i}e^{-i2\omega 
T-\epsilon})^{1/2}}
{1-e^{-i2\omega T-\epsilon}}\right),
}
\eqn\semifnl
$$
where we have renamed $2\omega T\epsilon$ as $\epsilon$
for brevity.  
For $T\neq k\pi/\omega $ with $k = 0, 1, 2, \ldots\, $,
we can take the limit 
$\epsilon\rightarrow +0$ safely to get
$$
\eqalign{
\left.S^{(1)} \right|_{T\neq k\pi/\omega }
&=
\sqrt{\frac{m\omega}{\hbar}}(|{y_f}{y_i}|)^{1/2}
\frac{1}{2i\sin(\omega T)}
\exp\left(\frac{i}{2}\frac{\cos(\omega T)}{\sin(\omega T)}
({y^2_f}+{y^2_i})\right)\cr
&\qquad\qquad\times \left[\Theta(y_fy_i) - \Theta(-y_fy_i)\right]\,
I_{{c_1}-1}\left(\frac{|{y_f}{y_i}|}{i\sin(\omega 
T)}\right).
}
\eqn\noncaust
$$
The contribution $S^{(2)}$ can be evaluated analogously 
and the result is exactly the same as $S^{(1)}$ except that $c_1$ is
now replaced by $c_2$ and the factor $[\Theta(y_fy_i) -
\Theta(-y_fy_i)]$ is removed in (\semifnl) or (\noncaust). 
Combining the two, for $T \ne k\pi/\omega$ we obtain the kernel
(\FeynmanKernelOne).

On the other hand, 
for $T=k\pi/\omega$ the kernel 
can be evaluated directly from (\inikernel).  From the 
energies (\inikernel) and 
the parity ${\psi^{(s)}_n} (-x) = (-1)^s{\psi^{(s)}_n} (x)$ of the eigenstates
(\SolutionTwo), we find
$$
\eqalign{
K({x_f},{t_f};{x_i},{t_i})
&=  \sum_{s=1, 2} e^{-ic_s k \pi} 
\sum^\infty_{n=0} {\psi^{(s)}_n} ({x_f})\,
({\psi^{(s)}_n}(x_i))^* \cr
&= \frac12\left(e^{-ic_1 k \pi} + e^{-ic_2 k \pi} \right)
\sum_{s=1, 2}\sum^\infty_{n=0} {\psi^{(s)}_n} ({x_f})\,
({\psi^{(s)}_n}(x_i))^* \cr
& \quad - \frac12\left(e^{-ic_1 k \pi} - e^{-ic_2 k \pi} \right)
\sum_{s=1, 2}\sum^\infty_{n=0} {\psi^{(s)}_n} (-{x_f})\,
({\psi^{(s)}_n}(x_i))^*.
}
\eqn\no
$$
Using the completeness of the
eigenstates and the relations $c_1 = 1 + a$, $c_2 = 1 - a$, we obtain
(\FeynmanKernelTwo).

\baselineskip= 15.5pt plus 1pt minus 1pt
\parskip=5pt plus 1pt minus 1pt
\tolerance 8000
\vfill\eject
\vfill\eject\immediate\closeout\reffile
\centerline{{\bf References}}\bigskip\frenchspacing%
\input refs.tmp\vfill\eject\nonfrenchspacing

\bye